\documentclass[aps,prl,twocolumn,showpacs,a4paper,superscriptaddress,showkeys]{revtex4}

\usepackage{dcolumn,amsmath,xspace}

\usepackage{graphicx}
\usepackage{dcolumn}
\usepackage{amsmath}
\usepackage{bm}

\usepackage{units}

\begin{document}
\title{Rotational stacking and its electronic effects on graphene films grown on 4H-SiC$(000\bar{1})$\\}
\author{J. Hass}
\affiliation{The Georgia Institute of Technology, Atlanta, Georgia
30332-0430, USA\\}
\author{F.Varchon}
\affiliation{Institut N\'{e}el/CNRS-UJF BP166, 38042 Grenoble
Cedex 9, France\\}
\author{J.~E. Mill\'{a}n-Otoya}
\affiliation{The Georgia Institute of Technology, Atlanta, Georgia
30332-0430, USA\\}
\author{M. Sprinkle}
\affiliation{The Georgia Institute of Technology, Atlanta, Georgia
30332-0430, USA\\}
\author{W.~A. de Heer}
\affiliation{The Georgia Institute of Technology, Atlanta, Georgia
30332-0430, USA\\}
\author{C. Berger}
\affiliation{The Georgia Institute of Technology, Atlanta, Georgia
30332-0430, USA\\}\affiliation{Institut N\'{e}el/CNRS-UJF BP166,
38042 Grenoble Cedex 9, France\\}
\author{P.~N. First}
\affiliation{The Georgia Institute of Technology, Atlanta, Georgia
30332-0430, USA\\}
\author{L.Magaud}
\affiliation{Institut N\'{e}el/CNRS-UJF BP166, 38042 Grenoble
Cedex 9, France\\}

\author{E.~H. Conrad}
\affiliation{The Georgia Institute
of Technology, Atlanta, Georgia 30332-0430, USA\\}

\begin{abstract}
We examine the stacking order of multilayer graphene grown on the
SiC$(000\bar{1})$ surface using low-energy electron diffraction
and surface X-ray diffraction. We show that the films contain a
high density of rotational stacking faults caused by three types
of rotated graphene: sheets rotated $30^\circ$ and $\pm
2.20^\circ$ relative to the SiC substrate. These angles are unique
because they correspond to commensurate phases of layered
graphene, both with itself and with the SiC substrate. {\it Ab
intio} calculations show that these rotational phases
electronically decouple adjacent graphene layers. The band
structure from graphene at fault boundaries displays linear energy
dispersion at the $K$-point (Dirac cones), nearly identical to
that of a single graphene sheet.
\end{abstract}
\vspace*{4ex}


\maketitle
\newpage

In the last few years an intriguing series of experiments suggests
that a new all-carbon paradigm for electronic circuits may be
possible \cite{Berger04,Novoselov05b,Zhang05c,Berger06}. In this
system graphene sheets are lithographically cut into ribbons to
produce gates and wires from a single material
\cite{Nakada96,edge2,Berger04}.  The promise of this new approach
to electronics rests on the ability to make large single or
multilayer graphene sheets on a substrate that preserves the
electronic properties of an isolated graphene sheet.

Both exfoliated graphene flakes \cite{Novoselov05b, Zhang05c} and
multilayer graphene grown on SiC \cite{Forbeaux00,Hass06} exhibit
2D transport properties characteristic of chiral Dirac electrons
expected for an isolated graphene sheet.  These include a Berry
phase of $\pi$ in the integer quantum Hall effect and suppressed
back-scattering \cite{Berger06,Novoselov05b,Wu07}. SiC-grown films
offer the most practical and scalable approach to 2D graphene
electronics, but the effect of the substrate and graphene stacking
must be clarified. Recent X-ray diffraction measurements and
electronic structure calculations showed that the
graphene-SiC$(000\bar{1})$ substrate interaction is weak, except
for a strongly-bound (non-conducting) buffer layer isolating
subsequent graphene layers from the SiC
\cite{Varchon_PRL_07,Hass_PRB_07}. However, the remaining paradox
is that normal {\it AB\ldots}stacked (Bernal) graphite breaks the
equivalency of {\it A} and {\it B} atoms in a graphene sheet
\cite{McCann_PRL_06,Latil_PRL_06} so that multilayer films should
not exhibit the graphene-like properties that are clearly observed
on 4H-SiC$(000\bar{1})$ substrates \cite{Berger06,Sadowski06}. As
we will show, in this case nature provides a new stacking sequence
that preserves the electronic symmetry of graphene.

In this Letter we present Low Energy Electron Diffraction (LEED)
and surface X-ray scattering data for multilayer graphene grown on
the 4H-SiC$(000\bar{1})$ carbon-terminated surface (C-face). We
show that graphene grows in three forms on this surface: layers
rotated $30^\circ$ ($R30$), or $\pm 2.20^\circ$ ($R2^{\pm }$) from
the SiC bulk $[10\bar{1}0]$ direction. In contrast, 4H-SiC$(0001)$
(Si-face) films orient only in the $R30$ phase (known as the
$(6\sqrt{3} \times 6\sqrt{3})R30^\circ$ reconstruction in SiC
coordinates) \cite{Forbeaux00}. X-ray diffraction on C-face films
confirms that all three rotated phases occur, causing a high
density of rotational fault boundaries between $R30$ and $R2^{\pm
}$ layers. {\it Ab initio} electronic calculations for this type
of stacking show that adjacent rotated planes become
electronically decoupled, preserving the Dirac dispersion at the
$K$-point. These results explain how, even in multilayer graphene,
the films maintain the electronic properties of an isolated
graphene sheet.

All substrates were 4H-SiC$(000\bar{1})$ prepared as previously
reported \cite{Hass06}. Once samples are graphitized, they remain
relatively inert, allowing them to be transported into the vacuum
chamber. The X-ray scattering experiments were performed at the
Advanced Photon Source, Argonne National Laboratory, on the 6IDB
and C-$\mu$CAT beam lines at $16.2~$keV photon energy. Samples
were mounted in a vacuum cryostat in the diffractometer.
Reciprocal space points are reported in the reciprocal lattice
units $(r.l.u.)$ of the standard graphite hexagonal reciprocal
lattice, ${\bf q}\! =\! (h{\bf a}^*,k{\bf b}^*,\ell{\bf c}^*)$,
where $|{\bf a}^*|\! =\! |{\bf b}^*|\! =\! 2\pi/(a\sqrt{3}/2)$ and
$|{\bf c}^*|\! =\! 2\pi/c$). The nominal lattice constants for
graphite are $a\! =\! 2.4589\text{\AA}$, $c\! =\! 6.674\text{\AA}$
\cite{Baskin_PR_55}.

\begin{figure}
\begin{center}
\includegraphics[width=6.5cm,clip]{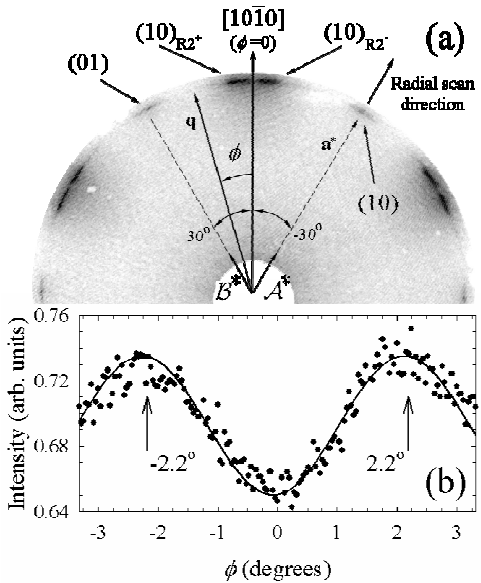}
\caption{(a) A LEED image acquired at 67.9eV from
4H-SiC(000$\bar{1}$) C-face with $\sim\! 10$ graphene layers
showing only graphite spots and diffuse rings.  The SiC
$[10\bar{1}0]$ direction is marked for reference. The SiC
$(6\sqrt{3}\! \times\! 6\sqrt{3})R30^\circ$ unit vectors
$\mathbf{\mathcal{A}}^*$ and $\mathbf{\mathcal{B}}^*$ are shown
for reference. (b) X-ray azimuthal scans of the diffuse graphite
ring around $\phi\! =\! 0.0$ and $|q|\! =\! |a^*|$.}\label{F:LEED}
\end{center}
\end{figure}

While it is known that graphene grows epitaxially only in the
$R30$ phase on the 4H-SiC$(0001)$ Si-face, multilayer graphene
grown on the C-face was thought to have a high degree of azimuthal
disorder because of streaking in LEED images \cite{Forbeaux00}.
However, a detailed look at the diffraction shows that the
rotational disorder is not random. This is demonstrated in the
LEED image  from a film with $\sim\! 10$ graphene layers
[Fig.~\ref{F:LEED}(a)]. The pattern has two characteristics; (i)
an oriented $R30$ film evidenced by graphene spots (rods) rotated
$\pm\! 30^\circ$ from the SiC $[10\bar{1}0]$ direction and (ii)
azimuthally diffuse rings centered at $0^\circ$ from the SiC
$[10\bar{1}0]$ direction. Note that the diffuse rings are not
continuous but split. This is seen more clearly in
Fig.~\ref{F:LEED}(b) that shows an X-ray azimuthal scan ($\phi$
scan) taken at the radial position of the graphite rod ($|q|=a^*$)
in Fig.~\ref{F:LEED}(a)) around $\phi=0.0^{\circ}$. The scan shows
diffuse intensity that is peaked at $\phi = \pm 2.2^\circ$.

The significance of the $\pm 2.2^\circ$ preferred rotation is
two-fold. First, graphene is nearly commensurate with the SiC
substrate on a length scale equal to a $(13\times 13)$ graphene
unit cell (rotated $30^\circ$ from SiC).  The $(13\!\times\! 13)$
cell is $\sim\! 0.14\%$ smaller than the SiC $(6\sqrt{3}\!
\times\! 6\sqrt{3})R30^\circ$ cell. What has not been recognized
is that there are two additional ways to orient a $(13\!\times\!
13)$ graphene sheet that give rise to a film commensurate with the
SiC $(6\sqrt{3}\! \times\! 6\sqrt{3})R30^\circ$ structure. They
can be calculated when the magnitude of the SiC $(6\sqrt{3} \times
6\sqrt{3})$ reciprocal lattice vector is nearly equal to the
graphene reciprocal lattice;
\begin{equation}
\left| n\mathbf{\mathcal{A}}^* +m\mathbf
{\mathcal{B}}^*\right|\approx \left|{\bf a}^*\right|.
\label{eq:Comm}
\end{equation}
$n$ and $m$ are integers. ${\bf \mathcal{A}}^*$ and ${\bf
\mathcal{B}}^*$ are the reciprocal lattice vectors of the
$(6\sqrt{3} \times 6\sqrt{3})R30^\circ$ structure (see
Fig.~\ref{F:LEED}(a)), $|{\bf \mathcal{A}}^*|\! =\! |{\bf
\mathcal{B}}^*|\! =\! |a^*_{\text{SiC}}|/(6\sqrt{3})$ with
$a^*_{SiC}\! =\! 2.3554\text{\AA}^{-1}$. The rotation angle of the
commensurate graphene relative to SiC can be calculated for
different $m$ and $n$'s:
\begin{equation}
\cos\phi = \frac{\sqrt{3}}{2}\frac{n+m}{\sqrt{n^2+m^2+nm}}.
\label{eq:Comm2}
\end{equation}

Equation (\ref{eq:Comm}) is satisfied when $(n,m)=(13,0)$, (8,7)
or (7,8). All three solutions give the SiC $(6\sqrt{3}\! \times\!
6\sqrt{3})R30^\circ$ cell but the first solution has graphene
rotated $30^\circ$ relative to SiC while the other two solutions
have graphene rotated $\pm 2.204^\circ$ relative to SiC. Graphene
grown on the Si-face of SiC only locks into the $30^\circ$
structure. However, as the diffraction in Fig.~\ref{F:LEED}(a)
clearly shows, all three phases appear on C-face grown multilayer
graphene. For future discussion we will index the two spots near
$\phi=0$ as the $(1,0,\ell)_{R2^+}$ and $(1,0,\ell)_{R2^-}$
graphene rods.  It is worth noting that, aside from the
$(1,0,\ell)_{R2^\pm}$ rods, other $(6\sqrt{3}\! \times\!
6\sqrt{3})R30^\circ$ reconstruction spots must be weak since they
are not seen in LEED. This is in contrast to Si-face grown
graphene where the additional spots indicate a complicated
interfacial graphene reconstruction. Apparently C-face graphene
has a weaker substrate interaction compared to Si-face graphene
that may explain why the additional rotated phases are specific to
C-face films.

\begin{figure}
\begin{center}
\includegraphics[width=5.4cm,clip]{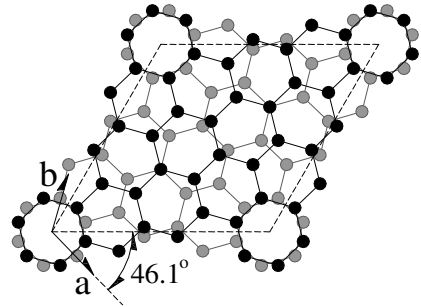}
\caption{Schematic $(\sqrt{13}\! \times\! \sqrt{13})R
46.1^{\circ}$ fault pair unit cell (dashed line). Dark circles are
$R30$ graphene atoms ({\bf a} and {\bf b} are graphene unit
vectors). Gray circles are graphene atoms in the $R2^+$ plane
below, rotated $32.204^\circ$ from the top plane.}
\label{F:Rotated_structure}
\end{center}
\end{figure}

The significance of these three phases is even more important if
we recognize that two graphene sheets can be rotated relative to
each other in a number of ways that make the two sheets
commensurate with each other \cite{Kolmogorov_PRB_05}. The lowest
energy commensurate rotation angles are precisely $\phi\! =\!
\cos^{-1}(11/13)$ or $\cos^{-1}(-11/13)\! -\! 120^\circ$) i.e,
$30\pm 2.204^\circ$ \cite{Kolmogorov_PRB_05}. This bi-layer
commensurate structure corresponds to a graphene $(\sqrt{13}\!
\times\! \sqrt{13})R(\pm 46.1^{\circ})$ cell that costs
3--5meV/atom more than {\it AB\ldots} stacking [a schematic of a
fault pair is shown in Fig.~\ref{F:Rotated_structure}]
\cite{Kolmogorov_PRB_05}.

\begin{figure}
\begin{center}
\includegraphics[width=7.0cm,clip]{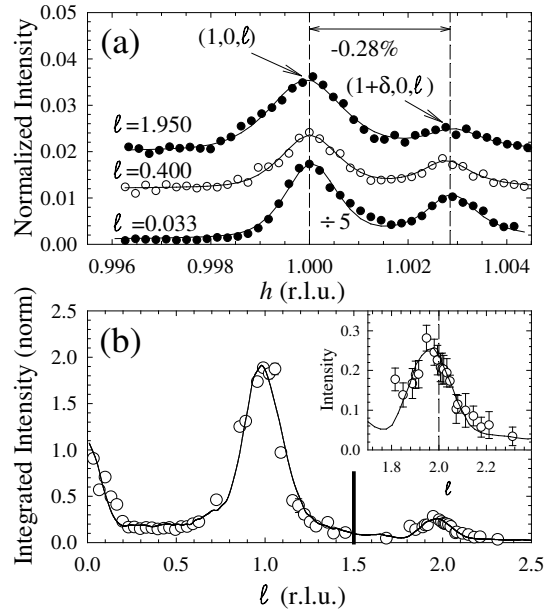}
\caption{(a) Radial ($h$) scans through the graphite
$(h=1,0,\ell)$ rod for different $\ell$ (see Fig.~\ref{F:LEED}).
Scans show two peaks corresponding to the normal graphene
$(1,0,\ell)$ rod and a compressed graphene $(1+\Delta h,0,\ell)$
rod. (b) Integrated intensity of the $(1,0,\ell)$ rod. Solid line
is a fit to a random rotational fault model. Vertical bar marks
the expected position of a peak from {\it ABC\ldots} faults. Inset
is an expanded view near $\ell =2$ showing a shift to smaller
$\ell$ (larger inter-layer spacing).} \label{F:Strain_peaks}
\end{center}
\end{figure}

While the observation of three rotational phases is interesting,
it is the stacking of these rotated planes that bears directly on
their electronic properties.  The $R30$ and $R2^\pm$ phases do not
exist as isolated domains. Instead all three rotations are present
in a multilayer graphene stack and lead to a high density of
$R30$/$R2^\pm$ fault pairs. The most direct evidence for this
comes from high-resolution X-ray diffraction.
Figure~\ref{F:Strain_peaks}(a) shows X-ray radial scans [see
Fig.~\ref{F:LEED}(a)] through the graphite $(1,0,\ell)$ rod for
different values of $\ell$. The two non-dispersing peaks
correspond to a normal graphene $(1,0,\ell)$ surface rod and the
other to a surface rod $(1+\Delta h,0,\ell)$ from graphene with a
compressed in-plane lattice constant. The peak separation
corresponds to a lattice compression of $\Delta a/a\!=\!\Delta
h/(1+\Delta h)\!=\! -0.28\pm 0.01\%$. Note that the compressed and
uncompressed rods widths are the same, meaning that the ordered
size of these two types of graphene sheets are similar. It is
important to realize that the compressed lattice is not a result
of epitaxial strain between the graphene and the SiC substrate,
which would increase the graphene in-plane lattice constant.

We can identify the compressed graphene as those sheets at the
$R30/R2^\pm$ fault boundary shown in
Fig.~\ref{F:Rotated_structure}. This conclusion is derived from
two key pieces of information found in the intensity modulation of
the $(1,0,\ell)$ rod [see Fig.~\ref{F:Strain_peaks}(b)]. First, in
normal graphite, the primary fault type is rhombohedral {\it
ABC\ldots} stacking. Faults of this type would produce a peak in
the $(1,0,\ell)$ at $\ell=1.5$ that is clearly not seen in the
data in Fig.~\ref{F:Strain_peaks}(b).

Second, {\it AA\ldots} and {\it ABC\ldots} faults are expected to
cause small inter-layer \emph{contractions} $\sim 0.2\%$
\cite{Charlier_Carbon_94}.  Instead, our X-ray results show a
large interplanar \emph{expansion} at the fault boundary. This is
shown in the inset of Fig.~\ref{F:Strain_peaks}(b) where the
experimental peak at $\ell=2$ is shifted to a slightly lower
value. This indicates a larger inter-layer spacing than bulk
graphite. We can estimate the interplanar expansion at the fault
using a model where random rotational faults are introduced with a
probability $\gamma$, and each $R30$/$R2^\pm$ boundary expands by
$\epsilon$.  A fit with $\gamma=0.38$ and
$\epsilon=0.06\text{\AA}$ is shown in
Fig.~\ref{F:Strain_peaks}(b).  This large expansion ($1.8\%$) is
characteristic of azimuthally disordered turbostratic graphite
with many rotational faults that cause significant interference of
$\pi^*$ states between rotated planes \cite{Baskin_PR_55}. Note
that this large interlayer expansion is also consistent with the
in-plane contraction of the fault. Graphite's  thermal expansion
is negative in-plane and positive out-of-plane
\cite{Barrera_JPCM_05}. A weaker inter-layer bond caused by the
large expansion at the fault allows the in-plane bonds to contract
\cite{Barrera_JPCM_05,Weinert_PRB_82}.

We expect significant changes in the electronic properties of the
graphene films when the rotated stacking fault density is as high
as we observe.  For a $R30/R2^{\pm }$ fault pair there are only 2
atoms/sheet out of 52 in the $(\sqrt{13}\! \times\!
\sqrt{13})R(\pm 46.1^{\circ})$ cell that are in high symmetry
positions.  This suggests weak interplanar interactions in the
fault pairs. To understand how the rotational stacking affects the
electronic properties of these films, we have performed an {\it ab
initio} Density Functional Theory (DFT) band structure calculation
for a $R30/R2^{\pm }$ free standing graphene fault pair. The
calculations are performed using the VASP ~\cite{vasp} code within
the generalized gradient approximation \!\cite{pw}.  Ultra soft
pseudopotentials \! \cite{uspp} are used with a plane wave basis
cutoff equal to 211 \!eV. All calculations are performed on the
same bi-layer commensurate cell, the $(\sqrt{13}\! \times\!
\sqrt{13})R(\pm 46.1^{\circ})$. This cell contains two graphene
sheets of 26 carbon atoms each, rotated $32.204^\circ$ relative to
each other [see Fig.~\ref{F:Rotated_structure}]. The empty space
width is equal to 24 \!\AA. The total energy of the rotated
stacking cell is 1.6meV/atom higher than an {\it AB..} cell. This
energy difference is slightly smaller than other estimates of this
structure \cite{Kolmogorov_PRB_05}.

For bilayers, the interlayer distance is fixed to 3.39\!\AA. As a
check, we have varied this distance and found no qualitative
change in the band structure results. This check was performed
since DFT is known to poorly describe van der Waals forces and
gives rise to theoretical graphene interlayer spacings
significantly larger than experimental values. However, we point
out that the C-short ultrasoft pseudopotential used here has been
extensively tested ~\cite{incze, Varchon_PRL_07} and was shown to
correctly describe the band structure of graphite . Integration
over the Brillouin zone is performed on a $30\times\! 30\times\!
1$ grid in the Monckhorst-Pack scheme to ensure convergence of the
Kohn-Sham eigenvalues. Due to the faulted cell symmetry, the
$K$-points of the graphene $1\times 1$ cell for the 2 layers of
the rotated stacking are translated to the $K$-point of the
$(\sqrt{13}\! \times\! \sqrt{13})R(\pm 46.1^{\circ})$ Brillouin
zone.

\begin{figure}
\begin{center}
\includegraphics[width=7.5cm,clip]{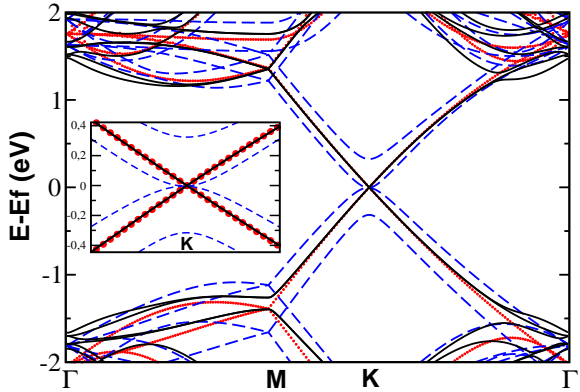}
\caption{Calculated band structure for three forms of graphene.
(i) isolated graphene sheet (solid line), (ii) {\it AB\ldots}
graphene bilayer (dashed line), and (iii) $R30/R2^+$ fault pair
(dots). Inset shows details of band structure at the $K$-point
showing no difference between the Dirac cone for an $R30/R2^+$
fault pair and a single graphene layer.} \label{F:Band_structure}
\end{center}
\end{figure}

The results of this calculation are shown in
Fig.~\ref{F:Band_structure} where we compare the band structure
for an isolated graphene sheet, a graphene bi-layer with Bernal
{\it AB\ldots} stacking, and a bilayer rotation fault pair with
the $(\sqrt{13}\! \times\! \sqrt{13})R(\pm 46.1^{\circ})$
structure (the $\Gamma$ $K$ $M$ direction shown is the
$(\sqrt{13}\! \times\! \sqrt{13})R\pm 46.1^{\circ}$ cell high
symmetry direction). The main differences in the electronic
structure among the three forms of graphene shows up in the
dispersion curves in the vicinity of the $K$-points.  The band
structure for an isolated graphene sheet shows the known linear
gapless dispersion (Dirac cone) of the $\pi$ bands at the
$K$-point.  The normal Bernal {\it AB\ldots} stacking of graphene
breaks the sublattice symmetry, giving rise to splitting of the
$\pi$ bands with a corresponding change to a parabolic shape and a
lower group velocity \cite{Latil_PRL_06}.  With the rotated fault,
the linear dispersion is recovered in the vicinity of the
$K$-points. This dispersion is identical to the graphene
dispersion (same Fermi velocity) and clearly shows that in the
rotated layers, the atoms in the {\it A} and {\it B} sublattices
are identical. This result also holds for infinite stacks: a
graphite-like system made of graphene sheets rotated alternately
by $0^\circ$ and $30\pm 2.204^\circ$ also shows a linear
dispersion near the $K$-point. This result is similar to the
continuum description of Santos et. al \cite{Santos_twisted
graphene} for two graphene sheets rotated by much smaller relative
angles.

In conclusion, we have shown that multilayer graphene grown on the
carbon terminated face of 4H-SiC does not grow as a simple {\it
AB\ldots} stacked film. Instead the graphene-SiC and
graphene-graphene commensurations produce a high density of
rotational faults where adjacent sheets are rotated $30^\circ \pm
2.204^\circ$ relative to each other. These rotational faults cause
adjacent graphene sheets to decouple electronically. The result is
that the band structure of faulted sheets is nearly identical to
isolated graphene. Specifically, the Dirac dispersion at the
$K$-point is preserved even though the film is composed of many
graphene sheets.  This may explain why magnetotransport
\cite{Berger06} and infrared magnetotransmission \cite{Sadowski06}
experiments on similar samples give results very similar to those
of an isolated graphene sheet.

We wish to acknowledge N. Wipf, G. Trambly de Laissardi\`{e}re and
D.Mayou for fruitful discussions about the energetics of rotated
graphene. This research was supported by the National Science
Foundation under Grant No. 0404084 and by Intel Research.  The
Advanced Photon Source is supported by the DOE Office of Basic
Energy Sciences, contract W-31-109-Eng-38. The $\mu$-CAT beam line
is supported through Ames Lab, operated for the US DOE under
Contract No.W-7405-Eng-82.

\end{document}